\documentclass[aps, twocolumn, superscriptaddress, floatfix, 
letterpaper, 
preprintnumbers]{revtex4}
\usepackage{graphicx}

\newcommand{\be}{\begin{eqnarray}}
\newcommand{\ee}{\end{eqnarray}}

\newcommand{\el}{\nonumber \hfill \\}
\newcommand{\Tr}{\mathrm{Tr}}
\newcommand{\nn}{\nonumber }
\begin{document}

%\preprint{SUNY-NTG-04/}

\title{Diquark and Pion Condensation in Random Matrix Models for
  two-color QCD}
\author {B. Klein} 
\affiliation{Institute for Theoretical Physics, University of
  Heidelberg, Philosophenweg 19, 69120 Heidelberg, Germany}
\author{D. Toublan}
\affiliation{Department of Physics, University of Illinois at 
Urbana-Champaign,
Urbana, IL 61801-3080}
\author{J.J.M. Verbaarschot}
\affiliation{Department of Physics and Astronomy, State University of New 
York at Stony Brook,
Stony Brook, NY 11794-3800}

\date{\today}
\begin{abstract}
We introduce a random matrix model with the symmetries of QCD with
two colors at nonzero isospin and baryon 
chemical potentials and temperature. 
We analyze its phase diagram  
 and find phases with condensation of pion and diquark states in 
addition to the phases with spontaneously broken chiral symmetries. 
In the limit of small chemical potentials and quark masses, we
reproduce the mean field results obtained from chiral Lagrangians.
As in the case of QCD with three colors, the presence of two
chemical potentials breaks the flavor symmetry and leads to phases that are
characterized by different behaviors of the chiral condensates for each
flavor. In particular, the phase diagram we obtain 
is                 
similar to QCD with three colors and three flavors of quarks of equal masses
at zero baryon chemical potential and nonzero isospin and strange
chemical potentials. 
A tricritical point of the superfluid transitions found in lattice
calculations and from an analysis in terms of chiral Lagrangians does
not appear in the random matrix model. 
Remarkably,  at fixed isospin chemical potential,
for the regions outside of the superfluid phases,
the phase diagram in the
temperature -- baryon chemical potential plane  for two colors and
three colors are qualitatively the same.

\end{abstract}
 
\pacs{}

\maketitle

\renewcommand{\theequation}{\arabic{section}.\arabic{equation}}
\setcounter{equation}{0}

\section{Introduction}
\label{nc2section-1}
\setcounter{equation}{0}
Recently, the phase diagram of QCD
at nonzero chemical potential for the baryon number has received
a great deal of attention. 
It has been found that a plethora of phase transitions are possible 
as a function of the baryon chemical potential, $\mu_B$, the isospin chemical
potential, $\mu_I$, and the temperature $T$. 
Among others, a superconducting phase
\cite{Alford:1997zt, Rapp:1997zu}, a pion condensate \cite{Migdal:jn,
  Son:2000xc} 
and a critical endpoint have been predicted \cite{Barducci:1989wi,
  Halasz:1998qr, Alford:1997zt} 
(see \cite{Rajagopal:2000wf} for a review). A systematic analysis of
the phase diagram is possible in the $\mu_B=0$ plane and, perturbatively,
for asymptotically large values of $\mu_B$ 
\cite{Son:1998uk}.
At finite, not asymptotically large baryon chemical potential, 
where the interactions between the 
quasi-particles that carry color charge are strong, the theoretical
basis of these predictions is rather weak.
So far, it has not been possible to confirm the
existence of the proposed phases in this region, 
neither from experimental results nor from lattice calculations. 
The reason for the failure of standard lattice
simulations is that in the presence of a chemical potential for baryons the
fermion determinant in the partition function becomes complex so that
a probabilistic interpretation of the partition function is not possible.

Thus, it is attractive to try to gain insight into the properties of such
condensed phases by studying QCD-like theories  
which avoid some of these complications. In this paper we study QCD with
two colors with baryons that are colorless diquark states. 
As in the case of
three colors, chiral symmetry is spontaneously broken and the theory
is confining at low temperatures and chemical potential. However, at
nonzero baryon number chemical potential,
two important simplifications occur in this theory.
First, the condensing diquark states are Goldstone bosons associated
with the spontaneous breaking of chiral symmetry with a critical chemical
potential equal to half the pion mass.
This makes it possible
to analyze the condensed phase by means of chiral Lagrangians
\cite{Kogut:1999iv, Kogut:2000ek, Splittorff:2001fy,
  Splittorff:2002xn, Splittorff:2000mm}.
Second, as a consequence of the pseudoreality of the $SU(2)$
gauge group, the fermion determinant remains real when a chemical potential is
introduced. Therefore, for an even number of flavors,
this theory can be simulated on the lattice without
encountering a sign problem. 

The phase diagram in two-color QCD has been studied using a variety of
methods. 
The appearance of a baryonic diquark condensate 
above a critical chemical potential of $\mu_{Bc} \simeq
m_\pi/2$ was first  observed in lattice simulations at strong
coupling \cite{Dagotto:1986gw, Dagotto:1986ms, Dagotto:xt}.     
More recently, the phase diagram was analyzed  in the framework of
chiral Lagrangians \cite{Kogut:1999iv, Kogut:2000ek,Splittorff:2001fy,
Splittorff:2002xn}, 
where it was found that the critical value of the
chemical potential is indeed given by the mass of the lightest state,
and that the transition is of second order at zero temperature.  
Lattice simulations confirm these results \cite{Aloisio:2000nr, 
  Aloisio:2000if, Aloisio:2000rb, Hands:1999md, Hands:2000ei,
  Hands:2000hi, Liu:2000in, Muroya:2001av, Muroya:2002ry,
  Kogut:2001if, Kogut:2001na, Kogut:2002tm, 
  Kogut:2002zg, Kogut:2002cm, Kogut:2003ju}.
In \cite{Kogut:2001if,Kogut:2002zg,Kogut:2002cm} 
it was observed that beyond a critical  value of the
baryon chemical potential and temperature, the transition to the
diquark condensation transition ceases to be of second order and becomes
a first order transition. This
implies the existence of a tricritical point.  In chiral
perturbation theory, such a tricritical point was found 
\cite{dunne,Splittorff:2001fy} at
values for the chemical potential and the temperature of the order of
the pion mass $m_\pi$.  

In physical applications, such as in neutron stars or
in heavy ion collisions, the chemical potentials for different quark
flavors are not necessarily equal. While the chemical potential for
baryon charge is the same for all quarks, an additional isospin
chemical potential parameterizes the difference between the chemical
potentials for two light quark flavors. Its effect on the phase
diagram  has been studied for $N_c=2$ in \cite{Splittorff:2000mm}
using effective chiral Lagrangians, and for $N_c=3$ in
\cite{Klein:2003fy} using a random matrix model, in
\cite{Toublan:2003tt, Frank:2003ve, Barducci:2004tt} by means of a
Nambu--Jona-Lasinio 
model, and in \cite{Barducci:2003un}
within the ladder approximation.  

In this article, we employ a random matrix model to study
QCD with two colors at finite 
chemical potentials for isospin and baryon density and finite
temperature. Originally, random matrix models have been introduced in  
QCD in order to describe correlations of low-lying eigenvalues of the
Dirac operator \cite{Shuryak:1992pi, Verbaarschot:1994qf}. These
random matrix models are equivalent to the static part of a chiral
Lagrangian which is uniquely determined by the symmetries of the
microscopic theory 
\cite{Osborn:1998qb, Damgaard:1998xy, Toublan:1999hi, Toublan:2000dn}. 
~From this connection, it has been
shown that they provide an exact analytical description of 
correlations in the low-lying Dirac spectrum. 
In addition to their use as exact analytical models for the spectrum,
random matrix models can also serve as schematic models for the phase
transitions in QCD. The first of these models was introduced in
\cite{Jackson:1995nf, Wettig:wi}
to describe the chiral restoration transition at
finite temperature. At finite baryon chemical potential, they have
been successfully used to explain the failure of the quenched
approximation in lattice QCD \cite{Stephanov:1996ki}, and to predict a
tricritical point for the chiral restoration transition at finite
density and temperature \cite{Halasz:1998qr}. The static part of the 
effective chiral Lagrangian for QCD with the coupling of charged
Goldstone bosons to a chemical potential has also been derived from
a chiral random matrix model \cite{Toublan:1999hx}. More recently, 
closely related models with a random gauge potential have been
considered for the superconducting phases \cite{Vanderheyden:1999xp,
  Vanderheyden:2000ti} and for diquark condensation for $N_c=2$ 
\cite{Vanderheyden:2001gx}. Though these models have additional
symmetries beyond those considered in this article, they lead to the
same spectral density as the corresponding chiral random matrix
ensembles \cite{Vanderheyden:2002gz}. 

The random matrix theory approach rests on the idea that it
is the matrix equivalent of a Landau-Ginzburg potential. Both theories
are based on the symmetries of the microscopic partition function. 
Indeed, as we will also see in this paper, it is possible to derive
a Landau-Ginzburg functional from the random matrix theory. 
The presence of a diquark and a pion condensation 
phase raises several interesting questions, which can be 
addressed by studying a random matrix model. In particular, we hope
to investigate the robustness of the tricritical point that has been
found in lattice simulations \cite{Kogut:2002cm}
and in chiral perturbation theory
\cite{Splittorff:2001fy}.
In general, we wish to identify which features of the phase diagram 
are generic and due to the underlying symmetries and which features
are model dependent.

This article is organized as follows. In section~\ref{nc2section-2}, we give a
short review of the symmetries and phase diagram of 
the partition function of QCD with two colors.
Our random matrix model is introduced in
section~\ref{nc2section-3}. We show
that it can be rewritten exactly in terms of the static part of
a chiral Lagrangian. With an ansatz
based on  the structure of the expected condensates we obtain an
effective potential for these condensates. Our main results are
presented in section~\ref{nc2section-5}, where we analyze the phase
diagram in the $\mu_B$-$\mu_I$ chemical potential plane.  
Concluding remarks are made in section~\ref{nc2section-6}.

\section{QCD with two colors}
\label{nc2section-2}
\setcounter{equation}{0}
The partition function for QCD with two colors 
is given by
\be
Z_{QCD} &=& \left \langle \prod_{f=1}^{N_f} \det(D+m_f + \mu_f 
\gamma_0)\right\rangle,
\label{ZQCDNc2article}
\ee
where the Euclidean Dirac operator is $D=\gamma_\mu
D_\mu=\gamma_\mu(\partial_\mu + iA_\mu)$, $\gamma_\mu$ are the
Euclidean $\gamma$-matrices, and $A_\mu$ is   
the $SU(2)$-valued gauge potential. For each of the $N_f$ quark flavors, the 
quark mass is given by $m_f$ and the chemical potential by $\mu_f$. The 
brackets denote an average over the Euclidean Yang-Mills action. In the case 
of only two flavors, the chemical potentials can be expressed in terms of 
isospin and baryon chemical potential, 
\be
\mu_B&=& \frac{1}{2} (\mu_1 + \mu_2) \el
\mu_I&=& \frac{1}{2} (\mu_1 - \mu_2).  
\ee

Due to the pseudoreality of the $SU(2)$ gauge group, the flavor
symmetry group is enlarged  to $SU(2N_f)$ from the 
usual $SU(N_f) \times SU(N_f)$ in
case of QCD with three or more colors.
In the vacuum, this symmetry
is spontaneously broken to $Sp(2N_f)$. Thus, in the chiral limit, 
one expects $N_f(2N_f-1) -1$ exactly massless
Goldstone bosons. Among these Goldstone bosons there are diquark
states which carry baryon charge, in addition to the mesonic pion
states which carry a charge with respect to isospin number. 
The mass term breaks the chiral flavor symmetry explicitly down to
$Sp(2N_f)$. A baryon chemical potential term by itself is invariant
under an $U(1)_B \times SU(N_f) \times SU(N_f)$ subgroup of the
original $SU(2N_f)$, and in the presence of a mass term this symmetry is
further broken down to $U(1)_B \times SU(N_f)$. 

The presence of an additional chemical potential coupling to light
bosonic states greatly increases the number of possible phases.
At low temperature and chemical potential we expect that chiral symmetry
is spontaneously broken. 
The phases with broken chiral symmetry are characterized by the
order parameters $\langle\bar u u\rangle$ and $\langle \bar d d \rangle$. 
As in the case of $N_c=3$ \cite{Klein:2003fy}, in the presence of two
different chemical potentials for the two flavors $u$ and $d$, there
is no reason to believe that $\langle\bar u u\rangle =\langle
\bar d d \rangle$, and in general, one expects that the chiral condensates
for both flavors are not equal. 
At $\mu_I \neq 0$, when the two chemical potentials are different, it cannot
be excluded that chiral symmetry
restoration will take place via two separate phase transitions.

In addition to the pion condensate $\rho=\frac{1}{2}(\langle \bar u
\gamma_5 d \rangle - \langle \bar d \gamma_5 u \rangle )$, already
known from the case $N_c=3$, which
appears above a critical isospin chemical potential, for QCD with two colors
we also have a
diquark condensate $\frac{1}{4} i ( \langle 
d^\dagger u^*\rangle - \langle u^\dagger  d^* \rangle + \langle d^T u
\rangle -  \langle u^T d \rangle ) $ above a critical baryon chemical
potential. ~From general arguments, one expects a phase transition 
when the chemical potential becomes
equal to the ratio of the mass and the  
charge for the lightest charged excitation.
Indeed, this has been shown explicitly by an analysis
in terms of chiral Lagrangians at low chemical
potential and temperature
\cite{Kogut:1999iv,Kogut:2000ek,Splittorff:2001fy, Splittorff:2002xn,
  Splittorff:2000mm}. 
It was also found that
the transition to a baryonic diquark condensation
phase or a pion condensation is of second order.

~From the symmetries of the partition function alone, it is possible to
predict some of the properties  of the phase diagram. 
As has been shown in \cite{Kogut:1999iv, Kogut:2000ek}, the
determinantal term in the partition function (\ref{ZQCDNc2article}) can be
rewritten in such a way that the enlarged flavor symmetry becomes
manifest. Using a chiral representation for the Euclidean
$\gamma$-matrices, in which
$\gamma_0\gamma_\mu=\mathrm{diag}(i\sigma_\mu, -i\sigma_\mu^\dagger)$
with the spin matrices $\sigma_\mu=(-i, \sigma_k)$, and the
pseudoreality property of the generators of the gauge group, 
$-\tau_2 \tau_a \tau_2=\tau_a^{*}$, 
the fermionic part of the Lagrangian with the 
Dirac operator can be rewritten in terms of a new spinor basis.
In the chiral limit $m_f \to 0$, but including a nonzero chemical potential,  
it becomes,
\be
\left(\begin{array}{c} \psi^*_L \\ \widetilde{\psi}^{*}_R\\\end{array}\right)^T
\left(\begin{array}{cc} 
i\sigma_\mu D_\mu +\mu_f & 0 \\
0 & i\sigma_\mu D_\mu - \mu_f \\
\end{array}\right)
\left(\begin{array}{c} \psi_L \\
    \widetilde{\psi}_R\\\end{array}\right),
\ee
where the spinors $\widetilde{\psi}_R=\sigma_2\tau_2 \psi^*_R$
transform in the same representation of the color group as the spinors
$\psi_L$. 
Clearly,
at $\mu_f=0$ and $m_f =0$, the flavor symmetry group is enlarged to
$SU(2N_f)$.
For two flavors, the fermion determinant in the partition
function factorizes into
\be
\lefteqn {
\det(m^2 -(iD_\mu\sigma_\mu + \mu_1)\sigma_2\tau_2
(i(D_\mu\sigma_\mu)^T - \mu_1)\sigma_2\tau_2) } \nn \\
&&\times 
\det(m^2 -(iD_\mu\sigma_\mu + \mu_2)\sigma_2\tau_2
(i(D_\mu\sigma_\mu)^T - \mu_2) \sigma_2\tau_2).\nn \\
\ee
The partition function is therefore invariant under the
interchanges $\mu_{f} \to -\mu_f$, $f=1, 2$, and $\mu_1
\leftrightarrow \mu_2$. 
 Translated into
the physical chemical potentials, the partition function is invariant
under the individual sign changes $\mu_I \to -\mu_I$ and $\mu_B \to
-\mu_B$, and under the interchange $\mu_I \leftrightarrow \mu_B$.    
The latter symmetry, which amounts to $\mu_2 \to
-\mu_2$, means the interchange of quarks and conjugate anti-quarks. Under
this transformation, a diquark condensate is transformed into a pion
condensate. 
Thus, overall one expects that the phase diagram is
symmetric about the lines 
$\mu_I^2=\mu_B^2$ and the lines $\mu_I=0$ and $\mu_B=0$. Therefore
from the symmetry of the QCD partition function we can restrict our
study of the phase diagram to one half quadrant in the
$\mu_B-\mu_I$-plane.

\section{Random Matrix Model}
\label{nc2section-3}
\setcounter{equation}{0}
We will introduce a random matrix model for QCD with
two colors and study its phase diagram at finite chemical potentials
for isospin and baryon 
number and finite temperature. 
The idea of the random matrix model is to replace the matrix elements of the 
Dirac operator by Gaussian distributed random variables, while respecting 
all the global symmetries of the QCD partition function. The dependence on 
temperature and chemical potentials will enter through external fields, 
which are also subject to symmetry constraints.

Besides the chiral and flavor symmetries, the Dirac operator can also have 
additional anti-unitary symmetries. For an $SU(2)$ gauge group, the Dirac 
operator satisfies the commutation relation 
\be
[ iD, C \tau_2 K] = 0, \;\;\;\; (C \tau_2 K)^2 = 1,
\ee
where $C=\gamma_2\gamma_4$ is the charge conjugation matrix, $\tau_2$ is a 
generator of the gauge group, and $K$ is the complex conjugation operator. 
Due to this anti-unitary symmetry, it is possible to find a basis in which 
all matrix elements of $D$ are real. Thus, the matrices in the
corresponding random matrix ensemble will have real
entries, 
which is denoted by the  Dyson index $\beta=1$.

In the absence of a temperature and chemical potential, the 
matrix elements of the Dirac operator are simply replaced by real random 
variables with a Gaussian distribution. The chemical potential breaks the 
flavor symmetry in the same way as in the QCD partition function.  
Just as the dependence on the chemical potentials, an effective temperature 
dependence enters through an external field \cite{Jackson:1995nf}. 
Because of 
the flavor symmetries and anti-unitary symmetries 
of the partition function the temperature field has to satisfy the 
following conditions
\begin{itemize}
\item The temperature term must be real.
\item It may not break the flavor symmetry.
\item Its eigenvalues are given by $\pm i n T$. In this article, we will 
consider only $n=1$. 
\end{itemize}
A real temperature term satisfying these constraints is given 
by the matrix 
\be
\omega(T)&=& \left(\begin{array}{cc} 0 & T \\ -T & 0 \end{array}\right).
\ee 
For vanishing chemical potential, such a temperature term leads to a second 
order phase transition \cite{Jackson:1995nf}. We wish to stress that
the temperature term here is   
different from the one for QCD with three colors. 
Naively taking the same temperature dependence as for the three color case 
breaks the flavor symmetry and leads to a very 
different phase diagram \cite{Vanderheyden:2001gx}.

The matrix representation for the Dirac operator in the random matrix model 
with two flavors is
\begin{widetext}
\be
D= \left(\begin{array}{cccc} 
m_1 & 0 &  W+ \omega(T)+ \mu_1 & 0\\
0 & m_2 & 0 &  W+ \omega(T)+ \mu_2 \\
-W^T -\omega(T)^T+\mu_1 & 0 & m_1 & 0\\
0 &  -W^T -\omega(T)^T+\mu_2 & 0 & m_2
 \end{array} \right),
\ee
\end{widetext}
where $W$ is a real $n \times n$ matrix. The probability distribution of its 
matrix elements is given by
\be
P(W)&=&\exp\left[ -\frac{n}{2}G^2 \Tr W^T W\right].
\ee
When we write the determinant as an integral over fermionic variables, the 
random matrix partition function becomes
\be 
Z_{RMT} &=& \int {\cal D}W \prod_f d \psi^f d \bar{\psi}^f P(W) \exp[ -
 \bar{\psi} D \psi ].\nn\\
\label{Zrmt}
\ee
We will also include source terms for a pion condensate, 
\be
\label{pionsource}
\lambda \frac{1}{2}\bar{\psi}\gamma_5 i \tau_2 \psi,
\ee
and for a diquark condensate, 
\be
\label{diquarksource}
j \frac{1}{4}( \psi^T \tau_2 \psi +  \bar{\psi}^T \tau_2 \bar{\psi}),
\ee
where the anti-symmetric Pauli matrix $\tau_2$ acts in flavor space. 

For a fixed number of flavors ($N_f=2$), the partition function is a 
function of  $\mu_1, \mu_2, T$ and $m_1, m_2, \lambda, j$. While the
masses act as source terms for the chiral condensates, they are also
parameters of the model. 

\section{Effective Partition Function}
\label{nc2section-4}
\setcounter{equation}{0}
Due to the unitary invariance of the random matrix partition 
function, it is possible to rewrite it in terms of effective degrees of 
freedom.  We will do this below, and then analyze the resulting effective 
partition function, first to lowest order in chiral perturbation theory
and then beyond this domain. 

Starting from the random matrix partition function (\ref{Zrmt}), we perform 
the Gaussian integrations over the elements of the random matrix. To 
decouple the resulting four-fermion terms, we perform a Hubbard-
Stratonovich-transformation and introduce an integration over the mesonic 
low-energy degrees of freedom. The integrations over the fermionic variables 
can then be performed exactly. 
The resulting partition function in terms of the new 
variables is
\be
Z^{\rm eff}&=& \int {\cal D} A \exp[-{\cal L}(A, A^\dag)], 
\ee
where
\be
{\cal L} &=&\frac{n}{2}G^2 \Tr A A^\dag -\frac{n}{4}\log \det Q'. 
\label{Leff}
\ee
The dependence on the chemical potentials and the temperature is contained 
in the $8 N_f \times 8 N_f$ matrix $Q'$
\begin{widetext}
\be
Q'&=&
\left(\begin{array}{cccc} 
A^\dagger + M  & 0 &\mu_I I_3 + \mu_B B & -T \\
0 & A^\dagger + M & T & \mu_I I_3 + \mu_B B \\
-(\mu_I I_3 + \mu_B B) & -T & A + M^\dagger & 0 \\
T & -(\mu_I I_3 + \mu_B B) & 0 & A+ M^\dagger \\
\end{array}\right).
\ee
\end{widetext}
The matrix $A$ is a complex, antisymmetric $2N_f \times 2N_f$ matrix. 
All the source terms are contained in the $2N_f \times 2N_f$ matrix $M$, 
which for $N_f=2$ is 
\be
\label{sourcematrix}
M= \left(\begin{array}{cccc}
  0 & m & -ij & \lambda \\
-m & 0 & \lambda & -ij\\
ij & -\lambda & 0 & m \\
-\lambda & ij & -m & 0 \end{array}\right ).
\ee
where we have  set $m_1=m_2=m$. ~From here on we will only consider quarks 
with equal mass.
The chemical potentials enter as the coefficients of the matrices for baryon 
and isospin charges, 
\be
B = \mathrm{diag}(1, -1, 1, -1),\;\;\;\; I_3 = \mathrm{diag}(1, -1, -1, 1). 
\ee   
This particular form of the charge matrices is due to the appearance of 
quarks and conjugate anti-quarks 
in the spinors transforming under the enlarged 
$SU(2N_f)$ symmetry, which have opposite charges with respect to the 
chemical potentials.

This effective Lagrangian can be simplified by a unitary transformation of the 
off-diagonal blocks, which leaves the value of the determinant unchanged. 
Exploiting the block structure, the effective Lagrangian (\ref{Leff})
can be expressed as
\be
{\cal L}&=&\frac{n}{2}G^2 \Tr A A^\dag -\frac{n}{4}\log \det Q_{+} Q_{-},
\label{Leff4}
\ee
where $Q_\pm$ are
the $ 4 N_f \times 4 N_f $ matrices
\be
Q_\pm\! = \!\! \left(\begin{array}{cc} 
A^\dag + M & \mu_B B + \mu_I I_3 \pm iT \\
-(\mu_B B + \mu_I I_3) \pm iT &  A + M^\dag
\end{array}\right)\!\! .                                     
\ee
The effective partition function with Lagrangian (\ref{Leff4}) 
is invariant under
\be
Q_\pm\to  U Q_\pm U^T, \quad \mu_2 \to -\mu_2, \quad \lambda \to j,
j \to -\lambda,
\label{symIB}
\ee
with 
\be
U =\left ( \begin{array}{cccc} 
1 & & & \\ & i\sigma_1 & & \\ & & 1 & \\ & & & -i\sigma_1 \end{array}
\right ) .
\ee 
This transformation thus corresponds to an interchange of $\mu_B$ and
$\mu_I$ and of the diquark condensate and the pion condensate.

We stress that this effective Lagrangian is an exact mapping of the
original random matrix partition function.

\subsection{Observables}
We wish to study the partition function and obtain the phase diagram with 
regards to four order parameters: the chiral condensates $\langle \bar u u 
\rangle$ and $\langle \bar d d \rangle$ for the two quark flavors, the pion 
condensate $\frac{1}{2}\left(\langle \bar u \gamma_5 d \rangle - \langle 
\bar d \gamma_5 u \rangle \right)$ and the diquark condensate $\frac{1}{4} i 
\left( \langle  d^\dagger u^*\rangle - \langle u^\dagger d^* \rangle + 
\langle d^T u \rangle -  \langle u^T d \rangle \right)$. We identify these 
observables by taking the derivatives of the partition function with respect 
to the sources. After partial integration the expressions for 
the chiral condensates for the two flavors are given by
\be
\langle \bar u u \rangle &=& \frac{1}{2n} \partial_{m_1} \log 
Z^{\mathrm{eff}}\el
&=& G^2 \frac{1}{4} \left\langle A_{21} -A_{12}+A_{21}^* - A_{12}^* 
\right\rangle  \el 
&\equiv& G^2 \sigma_1\el
\langle \bar d d \rangle &=& \frac{1}{2n} \partial_{m_2} \log 
Z^{\mathrm{eff}}\el
&=& G^2 \frac{1}{4} \left\langle A_{43} -A_{34}+A_{43}^* - A_{34}^* 
\right\rangle  \el
&\equiv& G^2 \sigma_2.
\ee
For the pion condensate we find in terms of expectation values for the 
elements of the matrix $A$
\be
\lefteqn{\frac{1}{2}(\langle \bar u \gamma_5 d \rangle - \langle \bar d 
\gamma_5 u \rangle )}\el
&=& \frac{1}{4n} \partial_{\lambda} \log Z^{\mathrm{eff}} 
\Big|_{\lambda=0}  \el
&=& G^2 \frac{1}{8} \langle A_{32} - A_{23} +A_{41} - A_{14}\el
& &   +A_{32}^* - A_{23}^* + A_{41}^* - A_{14}^* \rangle \el
&\equiv& G^2 \rho   
\ee
The diquark condensate is
\be
\lefteqn{\frac{1}{4} i ( \langle  d^\dagger u^*\rangle - \langle u^\dagger 
d^* \rangle + \langle d^T u \rangle -  \langle u^T d \rangle )}\el
&=&\frac{1}{4n} \partial_{j} \log Z^{\mathrm{eff}} \Big|_{j=0} \el
&=& G^2 \frac{1}{8} i \langle A_{13} - A_{31} + A_{24} - A_{42}\el 
& &    - A_{13}^* + A_{31}^* - A_{24}^* +A_{42}^* \rangle \el
&\equiv& G^2 \Delta.
\ee
In our convention, expectation values of the diagonal blocks 
correspond to the chiral condensates, and expectation values of 
the off-diagonal 
blocks correspond to the diquark and pion condensates. 

\subsection{Chiral Lagrangian}
The chiral Lagrangian
 is completely determined by the flavor symmetries of the 
microscopic theory and the pattern of the chiral symmetry breaking. Thus, it 
must be in agreement with the Lagrangian of chiral perturbation theory. For 
vanishing chemical potential and zero temperature, it has been shown 
that the random matrix partition function and the static part of the chiral 
Lagrangian in QCD are indeed equivalent 
\cite{Osborn:1998qb, Damgaard:1998xy, Toublan:1999hi}.

To show this equivalence in the present case, we will expand the
Lagrangian (\ref{Leff4}) 
in terms of the quark mass and the chemical potentials around the saddle 
point solution at $m=\mu_I=\mu_B=0$. In accordance with the usual power 
counting rules \cite{Kogut:2000ek}
we use $m={\cal O}(\epsilon^2)$ and $\mu_{I, B}={\cal 
O}(\epsilon)$ and expand to second order in $\epsilon$.
In contrast to the case $N_c=3$, the baryon chemical potential does not drop 
out of the effective Lagrangian. The baryon chemical potential couples to 
low energy degrees of freedom in form of diquark states, which carry a 
baryon charge. 

For vanishing mass and chemical potentials, the saddle point equation 
corresponding to (\ref{Leff4})
is given by
\be
\frac{1}{G^2} A &=& (A^\dagger A + T^2) A,
\ee
which has the nontrivial solution
\be
A&=&\frac{1}{G}\bar\sigma(T) \Sigma= \frac{1}{G}(1-G^2T^2)^{1/2} \Sigma,
\ee
where $\Sigma$ is any complex, antisymmetric, unitary matrix.
Expanding the effective Lagrangian (\ref{Leff4}) 
around this saddle point to order ${\cal O}(\epsilon^2)$, we find
\be
\lefteqn{{\cal L}(\Sigma) = \frac{n}{2} \left( c_0(\mu_B, \mu_I, T)\right.}\el
 & & - G\bar\sigma(T)\Tr(\Sigma M + \Sigma^\dagger M^\dagger) \el
 & & -G^2\bar\sigma^2(T)\Tr( (\Sigma^\dagger (\mu_I I_3 + \mu_B B) \Sigma 
(\mu_I I_3 + \mu_B B) )\el
 & & \left. +{ \cal O}(\epsilon^2)\right).
\ee
We note that the temperature does not break any flavor 
symmetries and has no effect 
on the matrix structure of $\Sigma$, as we expect for a correctly 
implemented temperature dependence. Its only effect is to change the overall 
magnitude of the condensates.
This result agrees with the static part of the chiral Lagrangian 
derived in \cite{Splittorff:2000mm}, which is (in our notation):
\be
{\cal L}_{\mathrm{stat}}(\Sigma) &=& -\frac{F^2 m_{\pi}^2}{2}\Tr(\Sigma M + 
\Sigma^\dagger M^\dagger) \el
& & \hspace{-3.5em} - \frac{F^2}{4} \Tr ( \Sigma^\dagger (\mu_I I_3 + \mu_B 
B) \Sigma (\mu_I I_3 + \mu_B B)). \nonumber \\   
\ee

This Lagrangian has been analyzed in detail in \cite{Splittorff:2000mm}. In the
remainder of this subsection we give 
a brief review of the main results of this paper. 
Let us first discuss the symmetries of the Lagrangian.
In the vacuum, the $SU(2N_f)$ symmetry is spontaneously broken
according to $SU(2N_f) \to Sp(2N_f)$, in the same way as it is broken
by the mass term. At nonzero baryon chemical potential,
 the symmetry is $SU(N_f) \times SU(N_f)$ and is  broken by the
diquark condensate to $Sp(N_f)\times Sp(N_f)$.
If in addition the mass is nonzero, the symmetry at $\mu \ne 0$
is $ U(1) \times SU(N_f)$ which is spontaneously broken by the
diquark condensate to $Sp(N_f)$ \cite{Kogut:2000ek}. 

The mass term and the terms for baryon and isospin chemical potential are 
individually minimized by $\Sigma$-matrices corresponding to a chiral 
condensate, a diquark condensate and a pion condensate, respectively. 
To minimize the Lagrangian, we use an ansatz consisting of a linear 
combination of these individual minima,
\be
\Sigma &=& \cos \alpha \Sigma_\sigma + \sin \alpha (\cos \eta \Sigma_\Delta 
+ \sin \eta \Sigma_\rho),  
\ee
which is again antisymmetric and unitary.
These matrices, which correspond to the source terms  
introduced in (\ref{pionsource}), (\ref{diquarksource}), and
(\ref{sourcematrix}), 
are given by 
\be
\Sigma_\sigma=\left(\begin{array}{cccc} 
0 & -1 & 0 & 0 \\
1 & 0 & 0 & 0 \\
0 & 0 & 0 & -1 \\
0 & 0 & 1 & 0 
\end{array} \right), \el
\Sigma_\Delta=\left(\begin{array}{cccc} 
0 & 0 & -i & 0 \\
0 & 0 & 0 & -i \\ 
i & 0 & 0 & 0 \\
0 & i & 0 & 0 
\end{array} \right), \el
\Sigma_\rho=\left(\begin{array}{cccc} 
0 & 0 & 0 & -1 \\
0 & 0 & -1 & 0 \\
0 & 1 & 0 & 0 \\
1 & 0 & 0 & 0 
\end{array} \right). 
\ee
Although the baryon chemical potential term is invariant under the
$U_B(1)$ rotations generated by $B$, this symmetry is spontaneously
broken by the diquark condensate, violating baryon number conservation
and giving rise to a massless Goldstone boson \cite{Kogut:2001if}. In
the same way, the pion condensate breaks the $I_3$ generated symmetry,
which also leads to a violation of isospin number conservation and a
Goldstone boson \cite{Son:2000by, Son:2000xc}.

\noindent
Using the ansatz for $\Sigma$, the effective Lagrangian (\ref{Leff4}) 
becomes
\be
{\cal L} &=& nN_f \left[ 1 + \log G^2 -2 m G \cos \alpha \right.\el
         & & + G^2 (\cos^2 \alpha (\mu_I^2 + \mu_B^2) -\sin^2 \alpha \cos (2 
\eta) (\mu_B^2 -\mu_I^2)) \el
         & & + \left.{\cal O}(\epsilon^2)\right]. 
\ee
When this is minimized with respect to $\eta$, for $\alpha \neq 0$ and 
$\mu_B^2 \neq \mu_I^2$, we find
\be
\eta= & 0 \;\;\;\mathrm{for} & \mu_B^2 > \mu_I^2\el
\eta= & \frac{\pi}{2} \;\;\;  \mathrm{for} & \mu_B^2 < \mu_I^2,
\ee
where the first case corresponds to the presence of a diquark condensate, 
and the second to that of a pion condensate. Thus, there is a first order 
transition between these two phases on the lines $\mu_B^2=\mu_I^2$. 
Minimizing now with respect to $\alpha$, we find 
\be
\cos \alpha   = & 1 & \mathrm{for} \;\;\;\mu_B^2 < \frac{m}{2G} 
\;\mathrm{and}\; \mu_I^2 < \frac{m}{2G} \el
\cos \alpha   = & \frac{m}{2G}\frac{1}{\mu_B^2} & \mathrm{for}\;\;\; \mu_B^2 
> \frac{m}{2G} \;\mathrm{and}\; \mu_B^2 > \mu_I^2 \el
\cos \alpha   = &\frac{m}{2G}\frac{1}{\mu_I^2} & \mathrm{for}\;\;\; \mu_I^2 
> \frac{m}{2G} \;\mathrm{and}\; \mu_B^2 < \mu_I^2.
\label{chptcond}
\ee
Since the critical chemical potential for the condensation of Goldstone 
bosons in our convention is half the pion mass, $\mu_{c} = 
m_\pi/2$, we can identify to this order in the expansion parameter
\be
m_\pi &=& \sqrt{\frac{2m}{G}}.
\ee

~From the analysis of the chiral Lagrangian we thus have obtained the following
picture of the phase diagram at zero temperature: 
below a chemical potential of half the pion 
mass, the chiral condensate is constant. Above this critical chemical 
potential, depending on the relative magnitude of isospin and baryon 
chemical potential, pion or diquark condensation sets in and the chiral 
condensate drops quickly. The phases of pion and diquark condensation phases 
are separated by first order transition lines at $\mu_I^2=\mu_B^2$.

\subsection{Effective Potential}
We now wish  to analyze the random matrix model to all orders in the mass and 
chemical potentials. To this end, we make an ansatz for the matrix $A$, and 
obtain an effective potential for the condensates in our model.
As in the case for three colors, the ansatz for $A$ is guided by two 
observations: ~From chiral perturbation theory, we know that above a critical 
chemical potential pion and diquark states will condense. We also expect 
that the presence of two flavor dependent chemical potentials, which 
explicitly breaks the flavor symmetry, will lead to a different behavior of 
the chiral condensates for the two flavors.  This leads to the
following ansatz  for $A$,
\be
A &=& \left( \begin{array}{cccc} 
0 & -\sigma_1 & -i\Delta &-\rho\\
\sigma_1 & 0 & -\rho & -i\Delta \\
i\Delta & \rho & 0 & -\sigma_2 \\
\rho & i\Delta & \sigma_2 &0 
\label{ansatz}
\end{array} \right).
\ee
With this ansatz, evaluating the determinant we obtain the effective
potential for the condensates: 
\begin{widetext}
\begin{eqnarray}
\frac{1}{n}{\cal L} 
 &=& G^2 (\sigma_1^2 +\sigma_2^2 +2 \Delta^2 + 2 \rho^2) -\frac 12 
{\rm Tr} \log Q_+ Q_-,
\label{pot}
\end{eqnarray}
where
\begin{eqnarray}
\det Q_{\pm} 
&=& \left \{
  [(\sigma_1+m+\mu_1\pm iT)(\sigma_2+m-\mu_2\mp
  iT)+\rho^2+\Delta^2]\right . \el 
 &&\hspace{2pt} \times \left .
  [(\sigma_1+m -\mu_1\pm iT)(\sigma_2+m+\mu_2 \mp iT)+\rho^2+\Delta^2] 
+ 4 \Delta^2 \mu_1\mu_2 \right\} \el
&& \hspace{0pt} \left\{
  [(\sigma_1+m-\mu_1\mp iT)(\sigma_2+m+\mu_2 \pm
  iT)+\rho^2+\Delta^2]\right. \el 
 &&\hspace{2pt} \times \left.
  [(\sigma_1+m +\mu_1\mp iT)(\sigma_2+m -\mu_2\pm iT)+\rho^2+\Delta^2] 
+ 4 \Delta^2 \mu_1\mu_2 \right\},
\end{eqnarray}
\end{widetext}
Although $\det Q_+$ and $\det Q_-$ are complex conjugate to each other by
construction, they turn out to be real themselves. 
With the ansatz (\ref{ansatz}) the symmetry (\ref{symIB}) corresponds
to the interchange (for $j = \lambda =0$)
\be
\Delta \to \rho, \quad \rho \to -\Delta, \quad \mu_2 \to -\mu_2.
\ee
The effective potential (\ref{pot}) is indeed invariant under this
symmetry, as well as the interchange of the flavor indices, and 
$\mu_f \rightarrow -\mu_f$ for both
$f=1,2$. These are the symmetries of the QCD partition function, as
shown in section~\ref{nc2section-2}. They allow us to restrict our study to one
half quadrant of the $\mu_B$-$\mu_I$--plane. The other sectors of
the phase diagram  are easily
obtained by applying these symmetries. 
The transition between the 
phases with diquark and pion condensation must necessarily be of first 
order and take place on the lines $\mu_I^2=\mu_B^2$.

\section{Phase Diagram}
\label{nc2section-5}
\setcounter{equation}{0}
In this section, we will  determine the complete phase diagram of the random 
matrix model in the space of  the chemical potentials, $\mu_B$ and $\mu_I$, 
and the temperature $T$. 
To this end, we solve the saddle point equations 
\be
\frac{\partial {\cal L}}{\partial \sigma_1}=0, \;\;\; 
\frac{\partial {\cal L}}{\partial \sigma_2}=0, \;\;\; 
\frac{\partial {\cal L}}{\partial \rho}=0, \;\;\; 
\frac{\partial {\cal L}}{\partial \Delta}=0 \;\;\;
\ee
and determine the expectation values of all the condensates in the saddle 
point approximation. For vanishing temperature,  we will obtain analytical 
solutions in the chiral limit as well as for finite quark mass. For nonzero 
temperature, we will present analytical results only in the chiral limit 
and study the potential numerically for finite quark mass.

\subsection{Chiral limit at zero temperature}
\setcounter{paragraph}{0}
First, we investigate the phase diagram in the $\mu_1$-$\mu_2$ chemical 
potential plane at temperature $T=0$. 
In the chiral limit, the Goldstone bosons arising from the spontaneous 
breaking of the chiral symmetry are exactly massless. 
Since all Goldstone bosons carry a charge that 
couples to one of the chemical potentials, this has a direct 
bearing on the phase diagram. At zero temperature, the chiral
condensate will be rotated completely into a bosonic condensate 
as soon as one of the chemical 
potentials becomes nonzero. 

\noindent
\paragraph{Chiral condensates and chiral restoration transition.}
In the phase where the bosonic condensates vanish, $\Delta=\rho=0$, 
the effective potential 
greatly simplifies and becomes a sum over the contributions of the chiral 
condensates for the two flavors,
\be
\frac1n {\cal L} &=& \sum_{f=1, 2} G^2 \sigma_f^2 -\frac{1}{2} \log
(\sigma_f^2-\mu_f^2)^2. 
\ee
Exactly as in the case of three color QCD, the saddle point equations
\be
\sigma_f [ G^2 (\sigma_f^2 - \mu_f^2) -1 ] =0, \;\;\; f = 1, 2, 
\ee
have the solutions
\be
\sigma_f  &=& 0, \; \; f=1,2, \\
\sigma_f^2&=& \frac{1}{G^2} +\mu_f^2,\;\; f=1, 2, 
\ee
which correspond to a phase with restored and broken chiral symmetry,
in this order. 
The contributions to the free energy for one flavor are
\be
\Omega_f(\mu_f) &=& - \log \mu_f^2 \\
\Omega_f(\mu_f) &=& 1 + \log G^2 +\mu_f^2 G^2,
\ee
for each of these phases, respectively.
The chiral symmetry restoration phase transitions take place at the 
chemical potential for which the two free energies coincide.
It is given by the solution of
\be
\label{limitmuccondition}
1+ \mu_f^2 G^2 + \log \mu_f^2 G^2 = 0. 
\ee
We note that this condition depends on only one of the two chemical 
potentials and is strictly constant in the other.  
The behavior of the two flavors in these phases is completely independent. 
This changes only in the presence of either a pion or diquark condensate, 
which couples both flavors.
As a result, the phases with one nonzero chiral condensate extend along the 
chemical potential axes. Only in the center of the phase diagram, where 
these strips overlap, could a phase exist for which both chiral condensates 
are nonzero.

\noindent\paragraph{Phases with bosonic condensates.}
At the point $\mu_1=\mu_2=0$, 
the 
solution of the saddle point equation is degenerate. It satisfies $\sigma_1 
\sigma_2 + \Delta^ 2 + \rho^2 = 1/G^2$, without a preference for a specific 
direction in the space of the condensates.  
Due to the absence of flavor symmetry breaking terms we expect 
$\sigma_1=\sigma_2$. This is the only point 
where all condensates can be simultaneously nonzero. 

In the chiral limit, the chiral condensates vanish at nonzero chemical 
potential in the presence of a bosonic condensate. As we argued above, the 
phases with diquark and pion condensation must be separated by a first order 
transition. This is confirmed by the saddle point equation: the only 
solution for which $\Delta \neq 0 $ \emph{and} $\rho \neq 0$ requires that 
either $\mu_1 = 0$ or $\mu_2 = 0$, which is equivalent to $\mu_I^2 = 
\mu_B^2$ and thus coincides with the expected first order transition line. 
Therefore, we always have either a pion or a diquark condensate, and the 
effective potential in these two cases is reduced to
\be
\frac{1}{n} {\cal L} &=& G^2 2 \rho^2 - \log(\rho^2 - \mu_1\mu_2)^2,\\
\frac{1}{n} {\cal L} &=& G^2 2 \Delta^2 - \log(\Delta^2 + \mu_1\mu_2)^2. 
\ee
The corresponding saddle point equations become
\be
G^2 \rho ( \rho^2 - \mu_1 \mu_2) - \rho &=& 0,\\
G^2 \Delta (\Delta^2 + \mu_1\mu_2) - \Delta &=& 0,
\ee
and have the respective solutions
\be
\rho &= 0, \;\;\; \rho^2 &=\frac{1}{G^2} + \mu_1\mu_2 \;\;\\
\Delta&=0, \;\;\; \Delta^2&=\frac{1}{G^2} - \mu_1 \mu_2. 
\ee
Second order phase transition lines are given by the conditions
$\mu_1\mu_2= -1/G^2$ and $\mu_1\mu_2= 1/G^2$. 
The free energies for the two solutions are
\be
\Omega_\rho &=& 2 ( 1 + \log G^2 + \mu_1\mu_2 G^2), \\
\Omega_\Delta&=& 2( 1 + \log G^2 - \mu_1 \mu_2 G^2).  
\ee
By comparing the free energies, we observe that in the quadrants with 
$\mu_1\mu_2 > 0$, the diquark condensate is nonzero,  
and that for $\mu_1\mu_2 < 
0$ the solution with a pion condensate has the lower free energy. Equating 
the free energies, we confirm again that first order transition lines are 
given by $\mu_1\mu_2=0$. 

\noindent\paragraph{Phase diagram.}
To find the complete phase diagram, we use the results we obtained directly 
for the second order phase transitions, and we compare the free energies of 
all phases to determine the first order transitions. We conclude that, 
for massless quarks, 
phases with condensed bosons are always favored above a phase where both
chiral condensates are non-vanishing. 
%%%%%%%%%%%%%%%%%%%%%%%%%%%%%%%%%%%%%%%%%%%%%%%%%%%%%%%%%%%%%%%%%%%%%%%%%%%
\begin{figure}
\hspace*{-0.4cm}
\includegraphics[scale=0.60, clip=true, angle=0, draft=false]{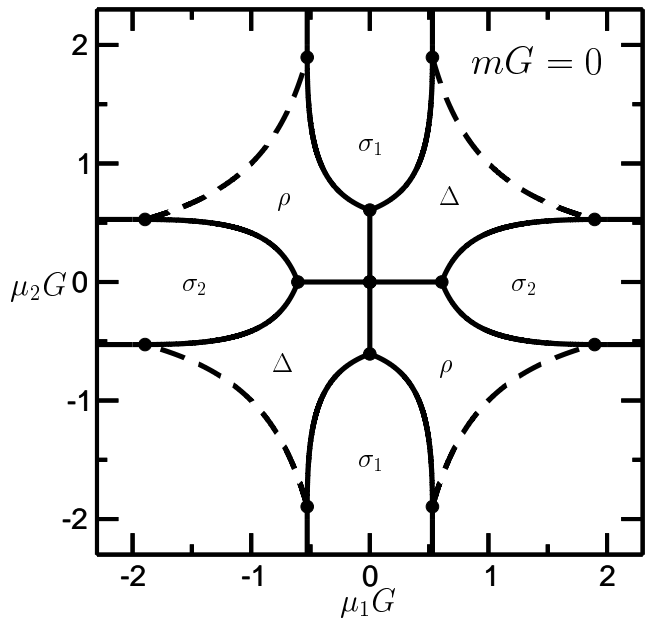}
\caption{\label{fig1} Phase diagram in the $\mu_1$-$\mu_2$-plane at $T=0$ 
and vanishing quark mass $mG=0$. First (second) order transitions are 
indicated by solid (dashed) lines. Phases are labeled by the nonvanishing 
condensates. All condensates vanish in the corner regions, where chiral 
symmetry is restored.}
\end{figure}
%%%%%%%%%%%%%%%%%%%%%%%%%%%%%%%%%%%%%%%%%%%%%%%%%%%%%%%%%%%%%%%%%%%%%%%%%%%
In addition to the first order transitions between the pion and diquark 
condensate phases, we find first order transitions from these to the phases 
where only one of the chiral condensates is different from zero. 

The phase diagram is shown in Fig.~\ref{fig1}.
At the origin, the solution to the saddle point equation is degenerate, and 
only the magnitude of the condensates is fixed. ~From the origin, a pion 
condensation phase extends into the quadrants where $\mu_1\mu_2 < 0$ 
($\mu_I^2 > \mu_B^2$), and a diquark condensation phase extends 
into the quadrants 
where $\mu_1 \mu_2 > 0$ ($\mu_I^2 < \mu_B^2$). The domain of these phases is 
limited by a line of second order transitions (see dashed curve in 
Fig.~\ref{fig1}), which 
in each quadrant is given by the solutions of 
\be
1 \pm G^2 \mu_1\mu_2 =0.  
\ee
These transitions can be interpreted as saturation
transitions which also occur in lattice QCD because of
the finite number of available fermionic states \cite{Kogut:2001na}.
Our saturation transition is somewhat different. Because we are working
in the thermodynamic limit, the number of available states is infinite
as well. However, since the level spacing of the relevant matrices
$\sim 1/n$, the support of the spectrum remains compact. It is the
non-analyticity that occurs because of the extreme edges of the spectrum
that gives rise to the saturation transition.
The phases with broken chiral symmetry and only {\it one} of the chiral 
condensate nonzero appear along both chemical potential axes. They are 
separated from the Goldstone condensation phases by first order transition 
lines.   

The phase diagram exhibits a fourfold symmetry, along both the
$\mu_B$- and the $\mu_I$- axis. This is in contrast to the twofold
symmetry with respect to the $\mu_B$-axis that we found in the phase
diagram of three color QCD, where only a pion condensate appears
\cite{Klein:2003fy}. Fundamentally, this is due to the flavor
symmetry in the two color case which is enlarged beyond the symmetry in
the three color case. 
However, the upper-left quadrant in the $\mu_B$-$\mu_I$--plane is
strictly identical in our random matrix model for both two and three
colors \cite{Klein:2003fy}.
The important point is that both
chemical potentials couple to Goldstone bosons of the theory.
Therefore, the two chemical potential terms in the present case are 
on equal footing. They lead to two different possible symmetry breaking 
patterns. For three colors, there is only one chemical potential term, since 
only the isospin chemical potential couples to the Goldstone modes, and thus 
there is only one way to break the flavor symmetry. 

\subsection{Nonzero quark mass at $T=0$}
\setcounter{paragraph}{0}

The phase diagram  in the presence of a finite quark differs from the one  
obtained in the chiral limit. 
As in the case of three colors, this is 
due to two effects: 
First, the chiral condensate becomes nonzero 
because of the explicit breaking of  chiral symmetry. Because of this, phase
transitions may turn into crossover transition.
Second, previously exactly massless 
Goldstone bosons become massive. This primarily affects 
the low energy regions of the phase diagram. In addition, the Goldstone 
condensation phases become qualitatively different and are now mixed phases 
with nonzero chiral condensates. In spite of this, at zero temperature we 
can still solve the saddle point equations exactly, and we will give the 
results below. 

\noindent
\paragraph{Chiral condensation phases.}
For a small quark mass, the behavior of the chiral condensate is still 
dominated by the spontaneous breaking of the chiral symmetry. 
Thus, a natural way to treat the chiral condensates in this case is to 
expand in the quark mass about the solutions for $m=0$.  
  
The effective potential that describes the phases without bosonic
condensates is given by
\be
\frac{1}{n}{\cal L} &=& \sum_{f=1, 2} G^2 \sigma_f^2 - \frac{1}{2} \log 
[(\sigma_f+m)^2 - \mu_f^2]^2. \nonumber \\
\ee
For each flavor we have to solve the saddle point equation
\be
G^2 \sigma_f [(\sigma_f+m)^2 - \mu_f^2] - (\sigma_f +m )=0.
\ee
As expected, the solution $\sigma_f=0$ no longer exists, and we find by 
expanding in $m$ (without assumptions about the magnitude of $\mu_f$) the 
new solution
\be
\sigma_f &=& - \frac{1}{1+\mu_f^2 G^2} m + {\cal O}(m^3).
\ee
Expanding in $mG$ in the phase with spontaneously broken chiral symmetry 
around the second solution with $G^2 \sigma_f = {\cal O}(G)$, we find the 
solutions
\be
G\sigma_f &=&\pm \sqrt{1+\mu_f^2G^2} 
+\Big(\frac{1}{2(1+\mu_f^2G^2)} -1\Big) mG \el
&& +{\cal O}(m^2G^2).
\ee 
The respective free energies for one flavor of these two phases with 
spontaneously broken and restored chiral symmetry are 
to lowest order in $m$ 
given by
\be
\Omega_f&=& 1 + \mu_f^2G^2 + \log G^2 \mp 2 mG \sqrt{1+\mu_f^2 G^2}, \\
\Omega_f&=& - \log \mu_f^2. 
\ee 
If these two phases are not separated by a Bose-con\-den\-sa\-tion phase and  
meet directly, the chiral restoration transition is of first order. The 
chemical potential at the transition is obtained by matching the free 
energies of the two phases. We find
\be
\mu^\prime_{f, c} G &=& \mu_{c}G\left(1 + \frac{mG}{1+\mu_c^2 G^2} + {\cal 
O} (m^2G^2) \right), 
\label{crshift}
\ee
where $\mu_{c}G$ is the critical chemical potential in the chiral
limit 
determined  from (\ref{limitmuccondition}). 
 The shift in the chiral restoration transition is proportional to the 
quark mass $m$.

\noindent
\paragraph{Goldstone boson condensation phases.}
For finite quark mass, the critical chemical potential for condensation of 
Goldstone bosons becomes nonzero due to their finite mass. The symmetry with 
regards to the chemical potentials  $\mu_I$ and $\mu_B$ remains unaffected. 
In particular, above the critical chemical potential, we still expect to see 
either a phase with diquark condensation or one with pion 
condensation depending on the relative magnitude of these two potentials, 
and a first order transition separating them. 
Since the results for both Goldstone condensation phases are related by this 
symmetry and are thus very similar, we will give explicit results for
one half quadrant only. 

In both phases, due to the finite quark mass, the chiral condensates do not 
vanish. Therefore, we have to solve three saddle point equations 
simultaneously, for the chiral condensates of both flavors and for either 
$\rho$ or $\Delta$. 
In the region in which $|\mu_I| > |\mu_B|$, we thus have $\Delta =0$.
The solutions of the saddle point equations are given by:
\be
\sigma_1 &=& -m + \frac{m}{2}\frac{1}{G^2}\frac{1}{\mu_I^2 -m^2} 
+ \frac12 m  (\mu_B + \mu_I) \frac{\mu_B}{\mu_I^2} \el
\sigma_2 &=& -m + \frac{m}{2}\frac{1}{G^2}\frac{1}{\mu_I^2 -m^2} 
+ \frac12 m (\mu_B -\mu_I) \frac{ \mu_B}{\mu_I^2} \el
\rho^2 + \sigma_1\sigma_2 &=& \frac{1}{G^2} +\mu_B^2-\mu_I^2 +m^2 -m^2 
\frac{\mu_B^2}{\mu_I^2}. 
\ee  
~From the relation between the chiral and the pion condensate, we find
the explicit result for $\rho$,
\be
\rho^2 &=& \mu_B^2 -\mu_I^2 + \mu_I^2 \frac{1}{G^2}\frac{1}{\mu_I^2 -m^2 } - 
\frac{m^2 }{4}\frac{1}{G^4}\frac{1}{(\mu_I^2 -m^2)^2} \el
        & & -m^2 \frac{\mu_B^2}{\mu_I^2}\left( \frac{1}{4 \mu_I^2}(\mu_B^2-
\mu_I^2) + \frac{1}{2}\frac{1}{G^2}\frac{1}{\mu_I^2 - m^2} \right).
\nonumber \\
\label{pc}   
\ee
The difference between the two chiral condensates depends on the ratio of 
$\mu_B$ and $\mu_I$: 
\be
\sigma_1-\sigma_2 &=& m \frac{\mu_B}{\mu_I}.
\ee
The free energy of this phase is given by
\be
\Omega_{\rho}(m, \mu_B, \mu_I) &=& 2 \left( 1 + \log G^2 + (\mu_B^2-
                \mu_I^2)G^2 +m^2 G^2 \right) \el
& & - m^2 G^2 \frac{ \mu_B^2}{\mu_I^2} 
       -\log \left(\frac{\mu_I^2}{\mu_I^2 -m^2 } \right). \nonumber \\
\ee

The condition $\rho^2=0$ 
describes a line of second order phase 
transitions at which the bosonic condensate vanishes. For finite quark mass, 
this condition has more than one solution. For $\mu_B=0$, the two solutions 
for $\mu_I^2$ are 
\be
\mu_{I\,c}^{2} G^2 &=& \frac{m G}{2} + \frac{5}{8} m^2 G^2 +{\cal 
O}(m^3 G^3)\\
\mu_{I\,c}^{2} G^2 &=& 1 + \frac{3}{4} m^2 G^2 + {\cal O}(m^4 
G^4). 
\ee 
We note that the solution for which $\mu_{I\, c}^2 G^2 = mG/2 + 
{\cal O}(m^2G^2)$ agrees to leading order with the result 
which was  
obtained from  
chiral perturbation theory \cite{Kogut:1999iv, Kogut:2000ek}. 
For chemical potentials below the critical value of $\mu_B$,
at which diquark condensation occurs, the pion mass is expected to be
independent from $\mu_B$ \cite{Splittorff:2000mm}. Thus, the critical value 
for $\mu_I$ should be exactly constant as
a function of $\mu_B$. In our random matrix model we find that
$\mu_{I\,c}$ is 
only
approximately constant as a function of $\mu_B$.
This is due to the fact that, in random
matrix models, the chemical potential affects the condensates already
below the threshold, even at zero temperature \cite{Stephanov:1996ki,
  Halasz:1998qr, Vanderheyden:2001gx}. 
In \cite{Halasz:1998qr, Vanderheyden:2001gx} it was 
thus argued that the random matrix
result should be interpreted as the critical contribution to the free
energy, on top of a noncritical part not contained in the model. 

For small  baryon chemical potential, small isospin 
chemical potential and quark mass,  
the chiral condensates in the pion condensation phase become
\be
\sigma_1=\sigma_2&=&\sigma=\frac{m}{2G^2 \mu_I^2} + {\cal O}(m)\\
\sigma^2 + \rho^2 &=& \frac{1}{G^2} +{\cal O}(m),
\ee
which agrees with the results (\ref{chptcond})
obtained in chiral perturbation theory. 
In this region of the phase diagram the effects of a finite quark mass are 
most significant. For large values of the chemical 
potentials both the pion and diquark condensates vanish 
for any value
of the quark mass. 
The second order transition to this phase is to leading order
independent of the quark mass.  

%%%%%%%%%%%%%%%%%%%%%%%%%%%%%%%%%%%%%%%%%%%%%%%%%%%%%%%%%%%%%%%%%%%%%%%%%%%
\begin{figure}
\hspace*{-0.4cm}\includegraphics[scale=0.60, clip=true, angle=0, 
draft=false]{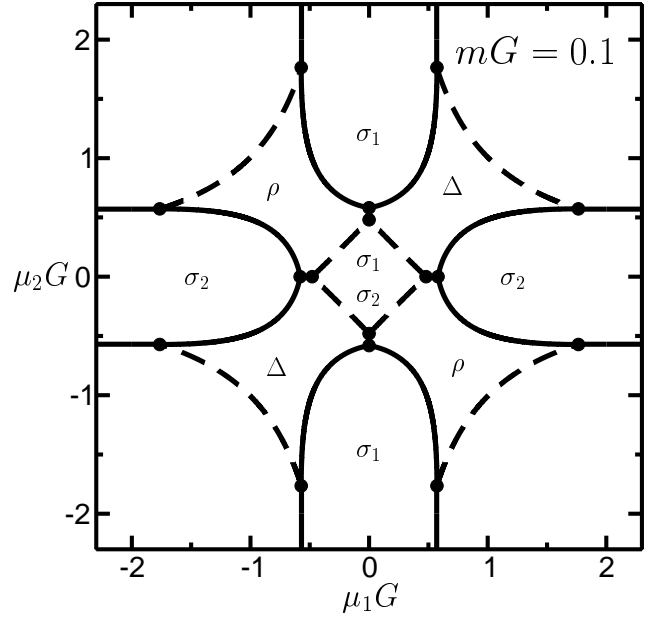}
\caption{\label{fig2} Phase diagram in the $\mu_1$-$\mu_2$-plane at $T=0$ 
for a finite quark mass  $mG=0.1$. First (second) order transitions are 
indicated by solid (dashed) lines. The different phases are labeled by the 
non-vanishing condensates or the condensates which have large
expectation values. The finite quark mass leads to a phase with chiral 
condensation only in the center of the phase diagram. In the corner regions, 
both the diquark condensate and the pion condensate vanish, 
and the chiral condensates are of 
${\cal O}(m)$. }
\end{figure}
%%%%%%%%%%%%%%%%%%%%%%%%%%%%%%%%%%%%%%%%%%%%%%%%%%%%%%%%%%%%%%%%%%%%%%%%%%%

%{$\bullet$ \bf Phase diagram}
\paragraph{Phase diagram.}

As in the chiral limit, in order 
to obtain the complete phase diagram we combine the 
results for the second order transitions with an analysis of the free 
energies of the different phases. The resulting phase diagram in the 
$\mu_1$-$\mu_2$--plane is shown in Fig.~\ref{fig2} for a quark mass of 
$mG=0.1$. The four-fold symmetry remains the same as in the case of $mG=0$. In 
particular, the transitions between the phases with the different Goldstone 
boson condensates are still of first order. The main difference from the 
chiral limit is the appearance of a phase in the center of the phase diagram 
in which the chiral condensates for both flavors are nonzero. This is a 
consequence of the finite critical chemical potential for the
condensation of bosons with a finite mass.  

A second effect of the finite quark masses is that 
the chiral restoration transitions 
outside the regions of Bose condensation are shifted slightly from their 
positions in the chiral limit. As can be seen 
from (\ref{crshift}), the  correction is of order $mG$. 
Thus, while changing the nature of the phase
transition considerably in the regions of low chemical potential 
in the center of the phase diagram, a finite quark mass has only a
small effect on the phase diagram at large chemical potential.     

%%%%%%%%%%%%%%%%%%%%%%%%%%%%%%%%%%%%%%%%%%%%%%%%%%%%%%%%%%%%%%%%%%%%%%%%%%%
\begin{figure}
\hspace*{-0.4cm}\includegraphics[scale=0.60, clip=true, angle=0, 
draft=false]{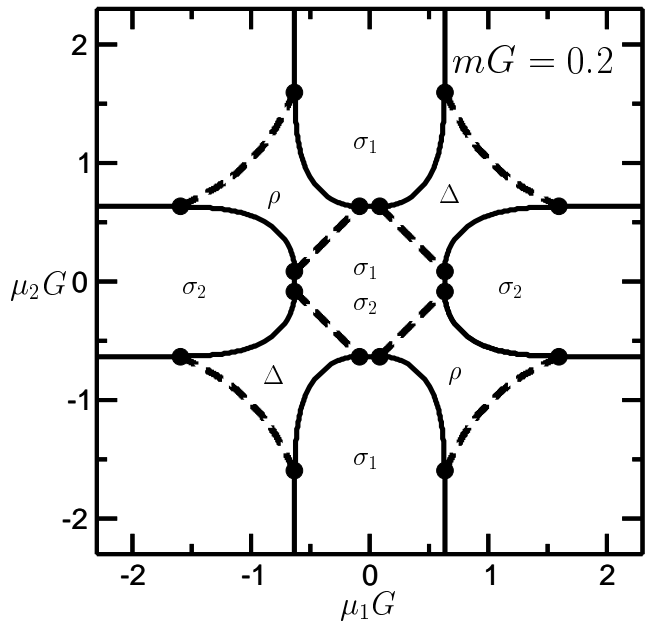}
\caption{\label{fig3} Phase diagram in the $\mu_1$-$\mu_2$-plane at $T=0$ 
for a finite quark mass  $mG=0.2$. First (second) order transitions are 
indicated by solid (dashed) lines. The different phases are labeled by the 
condensates which are non-vanishing or have a large expectation value. 
As described in the text, for a quark mass larger than $mG  \sim
0.184$ the pion and diquark condensation phases no longer
have a boundary in common. In contrast to the situation shown in
Fig.~\ref{fig2}, 
the chiral first order transition lines therefore appear also if one
of the chemical potentials $\mu_f$ is small.}
\end{figure}
%%%%%%%%%%%%%%%%%%%%%%%%%%%%%%%%%%%%%%%%%%%%%%%%%%%%%%%%%%%%%%%%%%%%%%%%%%%
For leading order in the mass expansion, we find that the first
order transition between the phases with $\rho\neq0$ and $\Delta\neq0$
disappears at the point $1+\log2mG=0$. Therefore, for masses larger than
$mG\sim0.184$, the pion and diquark condensation phases are not
contiguous anymore. 
In this case, if either $\mu_1$ or $\mu_2$ is kept small and the
chemical potential for the other quark flavor is increased, one
encounters the chiral restoration  
transition without passing through either the pion or the diquark
condensation phase.

\subsection{Nonzero $T$ at $m=0$}
\subsubsection{The $\mu_I=0$-- and $\mu_B=0$-planes}
\setcounter{paragraph}{0}
Since the presence of a temperature term does not change the 
symmetries, even at finite temperature 
the two phases with the different Goldstone condensates are 
still related by an interchange of the two chemical 
potentials. We can again treat both the pion and diquark condensation phases 
in parallel. In either the $\mu_I=0$ or the $\mu_B=0$ plane, the
presence of only one of these chemical potentials restores the symmetry between
the two quark flavors. Consequently, there is only one chiral condensate
$\sigma=\sigma_1=\sigma_2$. 
The results for the chiral phase transition
were first obtained for $\mu_I=0$ in \cite{Halasz:1998qr}.
The results in the plane $\mu_B=0$
coincide with  those for $N_c=3$ at nonzero 
isospin chemical potential \cite{Klein:2003fy} 
where the same effective potential
was found. Therefore, and mainly for completeness, we briefly
review the main results.
The result for the diquark condensate can be
obtained by the simple substitution $\mu_I \to \mu_B$ and $\rho \to
-\Delta$. 

In the chiral limit the critical chemical potential for pion condensation 
of this model 
is independent of the temperature. For any nonzero
value $|\mu_I| >0$, the system will be immediately in the pion
condensation phase (via a first order transition from the degenerate
point $\mu_I=0$). The effective potential for this phase is
\be
\frac{1}{n} {\cal L}&=& 2 G^2 \rho^2 -\log[ (\rho^2+\mu_I^2 +T^2)^2] ,
\ee
and the saddle point equations are solved by 
\be
\rho^2&=& \frac{1}{G^2} -\mu_I^2 -T^2, \\
\rho&=&0.
\ee
The phase diagram is now
simple: 
A second order transition
line is  given by  the circle $G^2 (\mu_I^2+T^2)=1$. The free
energy of this phase with $\sigma_1=\sigma_2 = 0$, given by
\be
\Omega_{\rho}&=& 2[ 1+\log G^2 - G^2(\mu_I^2+T^2)],
\ee
is always lower than that of the chiral broken phase, and thus in the
chiral limit, the chiral condensates do not appear in this 
plane 
(see the upper panel of Fig.~\ref{fig4}). Therefore, the 
tricritical point of the chiral restoration transition found in the
$\mu_B$-$T$--plane 
at $\mu_I=0$ 
for $N_c=3$ \cite{Halasz:1998qr} does not appear
at all in the $\mu_I$-$T$--plane
given by $\mu_B=0$. 
Using the symmetry of the phase diagram,
we see that, in contrast to the case $N_c=3$, for $N_c=2$
this tricritical point
is not present in the $\mu_B$-$T$--plane 
with $\mu_I=0$, 
either.  

\subsubsection{The complete $\mu_I$-$\mu_B$--plane at $T \neq 0$} 
\setcounter{paragraph}{0}
In the chiral limit, 
we can still solve
the saddle point equations analytically. 
Since the first order phase transition
between the pion and diquark condensation phases is not affected by the
temperature, we continue to have phases of pure diquark and pure pion
condensation which meet at the first order transition 
lines $\mu_1=0$ or $\mu_2=0$
(i.e. $\mu_I=\pm \mu_B$).
There are no phases in which a pion
condensate or a diquark condensate coexists with a chiral condensate
of any flavor.

\noindent\paragraph{Chiral condensation phases.}
The phase with only chiral
condensation occurs along strips parallel to the $\mu_1 =0$ or $\mu_2 =0$
axis (see Fig.~\ref{fig1}). In these phases only one of the chiral 
condensates is non-vanishing. 
The effective potential simplifies to a sum over 
two independent contributions of both flavors, 
\be
\lefteqn{\frac{1}{n} {\mathcal L} = \sum_{f=1, 2} \big\{ G^2 \sigma^2_f} \el
&& -\frac{1}{2}
\log\big[ (\sigma_f^2-\mu_f^2 + T^2)^2 + 4 \mu_f^2 T^2\big]\big\}, \el  
\ee
and coincides with the effective potential found for QCD with three
colors.
For each of the two flavors, the chiral condensate depends only
on the temperature and the chemical potential for that flavor. 
In the chiral limit, the magnitude of the chiral condensate is 
given by the solution of the saddle point equation
\be
\lefteqn{G^2 \sigma_f \Bigg[ \sigma_f^4-2 \left(\frac{1}{2G^2} +\mu_f^2 -T^2
    \right) \sigma_f^2} \el
&& + \frac{1}{G^2}(\mu_f^2 -T^2)+ (\mu_f^2 +T^2)^2\Bigg]= 0. 
\label{chspT}
\ee
Its  solutions are given by \cite{Halasz:1998qr}
\be 
\sigma_f &=& 0 \\
\sigma_f^2 &=& \frac{1}{2G^2} +\mu_f^2 -T^2 \pm \frac{1}{2G^2}
\sqrt{1-(4G^2\mu_f T)^2}, \el
\ee
and due to the solution $\sigma_f=0$ a first order
transition to the phase with restored chiral symmetry is possible. 
For any one flavor $f=1, 2$, this first order transition takes place
where the free energies of the two solutions, given by
\be
\Omega_f &=& -\log (\mu_f^2 +T^2) \\
\Omega_f &=& \frac{1}{2} + \log G^2 + G^2 (\mu_f^2 -T^2) \pm
\frac{1}{2} \sqrt{1-(4G^2 \mu_f T)^2} \el
 & & -\frac{1}{2} \log \left (\frac{1}{2}
\pm \frac{1}{2} \sqrt{1-(4G^2 \mu_f T)^2} \right),  
\ee
respectively, become equal.
The solution $\sigma_f \ne 0$ with the negative square root is not a global
minimum of the free energy, and is in fact never realized. We find
that the solution $\sigma_f \ne 0$ approaches the solution $\sigma_f = 0$
at the  second order phase transition
line in the $\mu_f$-$T$--plane given by 
\be
(\mu_f^2 -T^2) + G^2 (\mu_f^2+T^2)^2 =0. 
\ee
The tricritical point, where the phase transition line changes from
first to second order, appears where the coefficients of $\sigma_f$
and $\sigma_f^3$ in (\ref{chspT}) vanish. We find \cite{Halasz:1998qr}
\be 
\mu_{f,3}^2 G^2 &=& \frac{\sqrt{2}-1}{4},\quad f=1, 2 \el
T_3^2 G^2 &=& \frac{\sqrt{2}+1}{4}.
\label{nc2tricrit}
\ee
Since these results are independent of the chemical potential
for the other flavor, all transition lines are straight lines in the
$\mu_1$-$\mu_2$--plane at fixed temperature. In fact, the transition
lines for one flavor in this plane have to be perpendicular to the
corresponding lines of the other flavor. 

\noindent\paragraph{Goldstone boson condensation phases.} 
Once again relying on the symmetry of the 
effective potential (\ref{pot}), it is sufficient to analyze the 
pion condensate. Results for the diquark condensate
simply follow from  interchanging $\mu_I$ and $\mu_B$.  
The effective potential in the pion condensation phase at finite
temperature is given by  
\be
\lefteqn{\frac{1}{n}{\cal L} = 2 G^2 \rho^2 }\el
&&-\log[(\rho^2-\mu_B^2 +\mu_I^2+T^2)^2 + 4 \mu_B^2 T^2].
\ee
To explore the structure of the boundaries of the domain of this phase, 
we analyze the saddle point equation
\be
\lefteqn{G^2\rho \Big[ \rho^4 - 2\left(\frac{1}{2G^2} +\mu_B^2 -\mu_I^2 -T^2
  \right) \rho^2} \el
&&+(\mu_B^2-\mu_I^2 -T^2)^2 + 4 \mu_B^2 T^2
  +\frac{1}{G^2}(\mu_B^2-\mu_I^2-T^2)\Big] =0.\el
\label{nc2sprhot}
\ee  
Besides the trivial solution $\rho=0$, the saddle point equation has
the solutions
\be
\rho^2&=& \frac{1}{2G^2} +\mu_B^2 -\mu_I^2 -T^2 \pm \frac{1}{2G^2}
\sqrt{1-(4G^2\mu_B T)^2},\el  
\ee
of which only the one with the positive sign in front of the
square root is a minimum of the free energy. 
A second order transition line is given by the condition that this solution
coincides with the trivial solution,
\be
\lefteqn{(\mu_B^2-\mu_I^2-T^2)^2 + 4 \mu_B^2 T^2} \el
&&+ \frac{1}{G^2} (\mu_B^2-\mu_I^2-T^2)=0.
\label{nc2rhosectran}
\ee
There is always the
possibility of a first order transition to a phase with $\rho=0$.
Indeed,  for a certain range of
parameters, where the coefficients of
both $\rho^3$ and $\rho$ in eq. (\ref{nc2sprhot}) vanish, there exists a
tricritical point at which the transition changes order. But when the
complete phase diagram is considered, including all other condensates,
the solutions that lead to this tricritical point do not appear: 
at this point the free energy
of one of the chiral condensation phases is lower than that of the
pion condensation phase, and a first order transition 
takes place before the point is reached.
The free energy in the pion condensation phase is given by
\be
\Omega_{\rho}&=& 2\bigg\{ \frac{1}{2} + \log G^2 + G^2(\mu_B^2 -
  \mu_I^2-T^2)\el
&&  +\frac{1}{2} \sqrt{1-(4 G^2 \mu_B T)^2} \el
&&-\frac{1}{2}\log\bigg[\frac{1}{2}  +\frac{1}{2} \sqrt{1-(4 G^2 \mu_B
      T)^2} \bigg] \bigg\}, \el
\ee
and we determine the complete phase diagram by comparing the free
energies of the different phases. Using the symmetry (\ref{symIB}),
all the above results also apply to  the diquark
condensate.

\noindent\paragraph{Phase diagram.}
At finite temperature, the phase diagram in the $\mu_1$-$\mu_2$ plane
still has the fourfold symmetry as seen at $T=0$ in 
Fig.~\ref{fig1}. 
As can be seen from the second
order transition condition (\ref{nc2rhosectran}), the boundaries of
the regions with diquark condensation or pion condensation 
move gradually towards the origin  $\mu_B=\mu_I=0$ with increasing
temperature, and shrink rapidly to a point when $TG$ approaches $TG=1$. 
The regions with a nonzero chiral condensate also shrink gradually
with increasing temperature. When the temperature reaches the
tricritical point  given by
eq. (\ref{nc2tricrit}), the transition to the chirally restored region
becomes second order. As the temperature approaches $TG=1$ this region
rapidly collapses onto the axes $\mu_1 =0$ and $\mu_2=0$. 
All condensates vanish and the symmetries are completely restored for
$TG \ge 1$. 

\subsection{Finite quark mass at nonzero $T$}

In general, for $m\neq 0$ the  saddle point equations cannot be solved
analytically when $\mu_I$, $\mu_B$ and $T$ are all different from zero. 
However, as outlined in the previous section, the symmetries of the
partition function under an interchange of the chemical potentials for
the two flavors, or a change in their sign, are unaffected by the
presence of the temperature term. Thus, the four symmetry axes $\mu_I=0$,
$\mu_B=0$ and $\mu_I=\pm
\mu_B$ remain at finite 
temperature, and the first order transitions between the condensates
$\Delta$ and $\rho$ at  $\mu_I=\pm \mu_B$ remain. 
The structure of the phase diagram in this case can be deduced from
the analytical results at $T=0$ and from the planes where one of the
chemical potentials vanishes.

\subsubsection{The $\mu_B=0$-- plane}
\setcounter{paragraph}{0}

%%%%%%%%%%%%%%%%%%%%%%%%%%%%%%%%%%%%%%%%%%%%%%%%%%%%%%%%%%%%%%%%%%%%%%%%%%%
\begin{figure}[h!]
\hspace*{-0.9cm}\includegraphics[scale=0.85, clip=true, angle=0,
  draft=false]{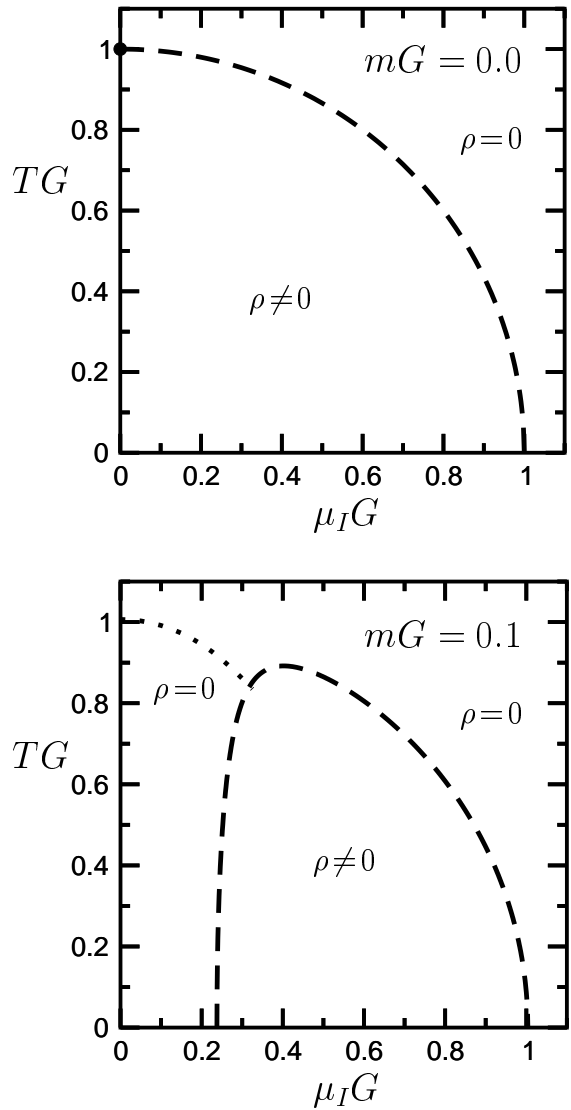}
\caption{\label{fig4} Phase diagram in $\mu_I$ and $T$ in the
  $\mu_B=0$-plane, both for the chiral limit (top)  
and a finite quark mass $mG=0.1$ (bottom). First (second) order
  transitions are  
indicated by solid (dashed) lines, the dotted lines in the case of
  finite quark mass mark the crossover.} 
\end{figure}
%%%%%%%%%%%%%%%%%%%%%%%%%%%%%%%%%%%%%%%%%%%%%%%%%%%%%%%%%%%%%%%%%%%%%%%%%%%

%%%%%%%%%%%%%%%%%%%%%%%%%%%%%%%%%%%%%%%%%%%%%%%%%%%%%%%%%%%%%%%%%%%%%%%%%%%
\begin{figure*}[h!]
\hspace*{-0.9cm}\includegraphics[scale=0.94, clip=true, angle=0,
  draft=false]{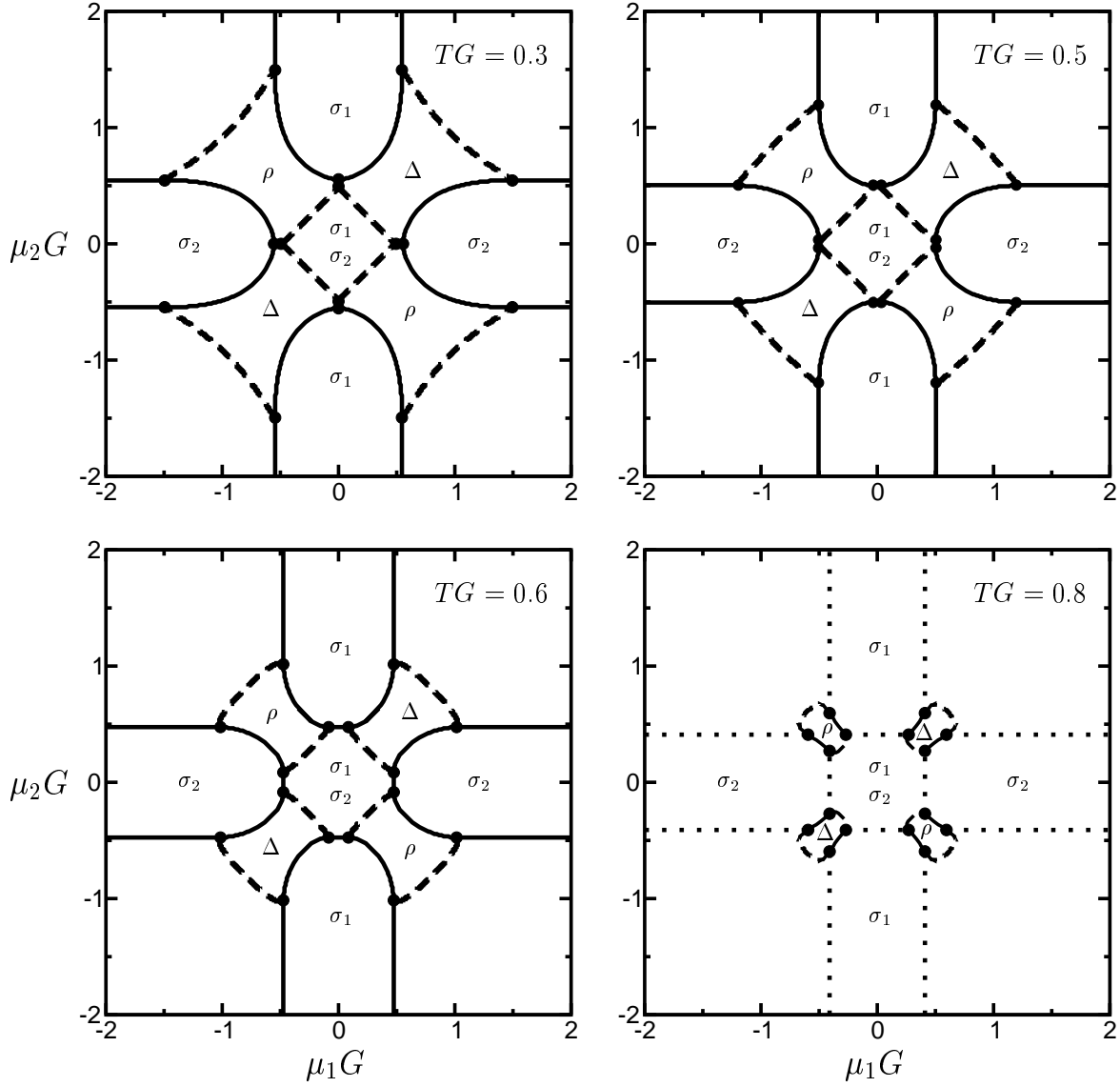} 
\caption{\label{fig5} Phase diagram in the $\mu_1$-$\mu_2$--plane at
  fixed values of $TG$ 
and for a finite quark mass of $mG=0.1$. Values for the temperature $TG$ are
  given in the panels. First (second) order transitions are 
indicated by solid (dashed) lines, the dotted lines indicate a
  crossover. Phases are labeled by the non-vanishing 
condensates, resp. the condensates with large expectation values. In
  the corner regions, the pion and diquark condensates vanish, and
  both chiral condensates become small.} 
\end{figure*}
%%%%%%%%%%%%%%%%%%%%%%%%%%%%%%%%%%%%%%%%%%%%%%%%%%%%%%%%%%%%%%%%%%%%%%%%%%%

Although an analytical solution of the saddle point equations is not
possible for both chemical potentials and the temperature nonzero, in
order to have a complete picture of the phase diagram it is
instructive to consider the plane $\mu_B=0$. Here, of course, the
chiral condensates for both flavors coincide again.
This region of the phase diagram has been studied at zero temperature in
\cite{Son:2000xc} and \cite{Splittorff:2000mm} by means of a
chiral Lagrangian for small values 
of $\mu_I$.
Finite temperature effects have been included in a one-loop
calculation in chiral perturbation theory in \cite{Splittorff:2002xn} and
in the random gauge model of \cite{Vanderheyden:2001gx}. 

%%%%%%%%%%%%%%%%%%%%%%%%%%%%%%%%%%%%%%%%%%%%%%%%%%%%%%%%%%%%%%%%%%%%%%%%%%%
\begin{figure*}
\hspace*{-0.9cm}\includegraphics[scale=0.92, clip=true, angle=0,
  draft=false]{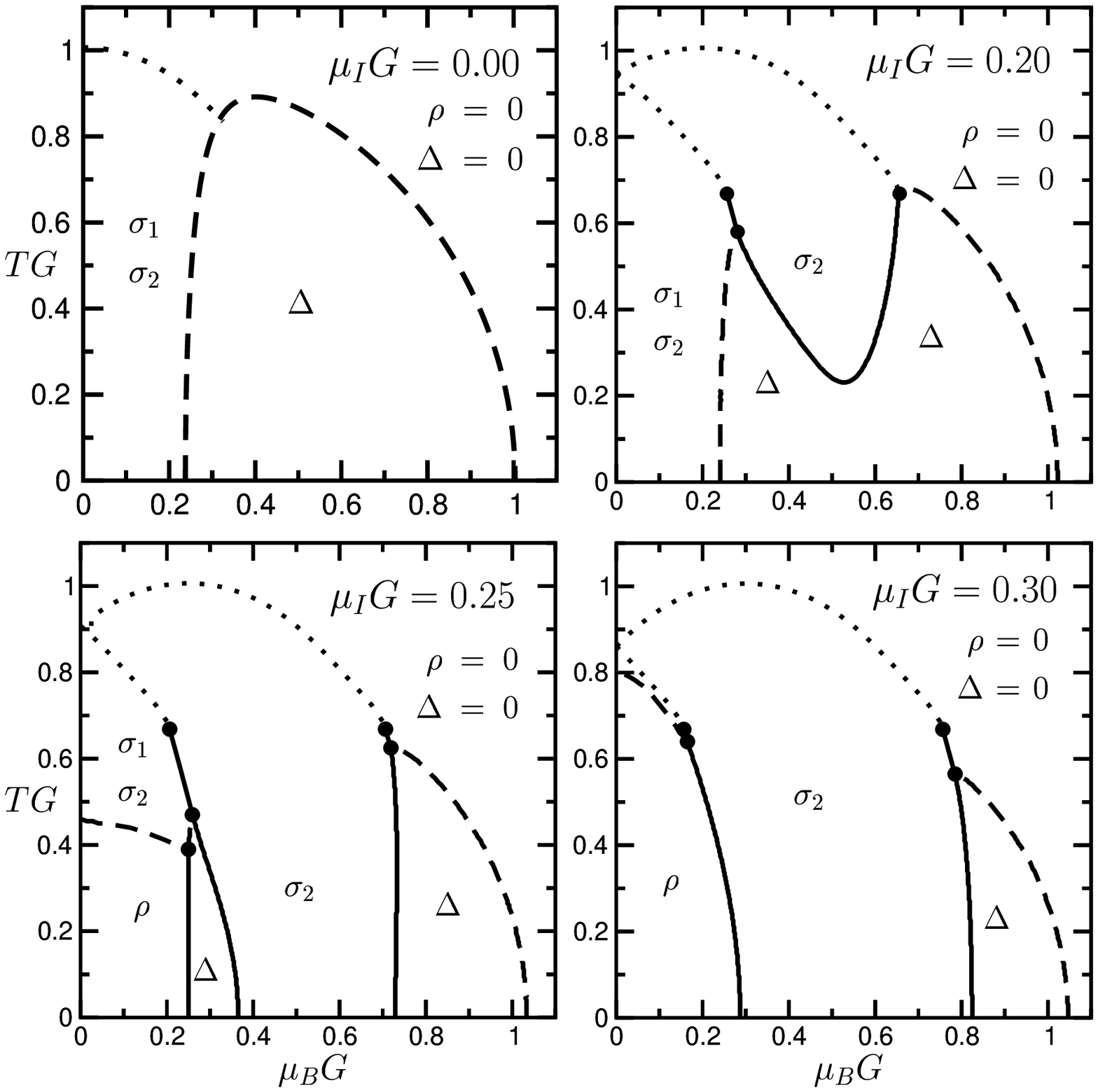} 
\caption{\label{fig6} Phase diagram in the $\mu_B$-$T$--plane at
  fixed values of the isospin chemical potential $\mu_I G$ 
for a finite quark mass of $mG=0.1$. Values for $\mu_I G$ are
  given in the panels. First (second) order transitions are 
indicated by solid (dashed) lines, the dotted lines indicate a
  crossover. Phases are labeled by the non-vanishing 
condensates, or the condensates with large expectation values.}
\end{figure*}
%%%%%%%%%%%%%%%%%%%%%%%%%%%%%%%%%%%%%%%%%%%%%%%%%%%%%%%%%%%%%%%%%%%%%%%%%%%

\paragraph{Chiral condensate.}
Below the critical value of $\mu_I$, chiral symmetry is broken by 
the chiral condensate, and at high temperature it is restored after a
crossover transition. 
For $m \neq 0$, it is most transparent to solve the saddle point
equations for the chiral condensate by expanding in the quark
mass $m$.
In the broken phase, the chiral condensate $\sigma=\sigma_1=\sigma_2$
is given by 
\be
\sigma=\sigma_0 + m \frac{1}{2\sigma_0^2}
\frac{\sigma_0^2-\mu_I^2+T^2-8 G^2 T^2 \mu_I^2}{1-(4G^2\mu_I T)^2} -m
+{\cal O}(m^2)  
\nonumber \\
\ee 
where $\sigma_0$ is the solution of the saddle point equation for
$m=0$, 
\be
\sigma_0^2&=&\frac{1}{2G^2} +\mu_I^2-T^2 \el
&&+\frac{1}{2G^2} \sqrt{1-(4G^2\mu_I T)^2}.
\ee
With increasing temperature, the chiral condensate decreases. In the
high temperature phase, the chiral
condensate can be expanded about the trivial solution (5.39) as
\be
\sigma=-m\frac{\mu_I^2-T^2}{G^2 (\mu_I^2+T^2)^2
  +\mu_I^2-T^2} + {\cal O}(m^3).
\ee

\paragraph{Pion condensate.}
As in the case $T=0$, for finite quark mass the chiral condensate does not
vanish in the pion condensation phase. In this mixed phase, we 
can solve the saddle point
equations  analytically. The solution given by
\be
\sigma&=&-m+\frac{m}{2G^2} \frac{1}{\mu_I^2-m^2}\\
\sigma^2+\rho^2&=& \frac{1}{G^2} +m^2 -\mu_I^2 -T^2. 
\label{nc2rotate}
\ee
was first  obtained for diquark condensation \cite{Vanderheyden:2001gx}.
We notice that the chiral condensate is completely independent of the
temperature, and only the total magnitude $\sigma^2+\rho^2$ of the
condensates decreases with increasing temperature. At zero temperature
these equations describe the rotation of a chiral condensate into
a pion condensate for increasing $\mu_I$.
The explicit solution for the pion condensate obtained from
(\ref{nc2rotate}) is given by
\be
\rho^2&=& \frac{1}{G^2} \frac{\mu_I^2}{\mu_I^2-m^2} \el
&& -\frac{1}{4G^4}
\frac{m^2}{(\mu_I^2-m^2)^2} -\mu_I^2 -T^2.
\ee
The phase with $\rho \ne 0$ is 
bounded by a second order line at which $\rho$ vanishes:
\be
\lefteqn{\mu_I^2(\mu_I^2-m^2) -\frac{1}{4}m^2} \el
&&-(\mu_I^2+T^2)(\mu_I^2-m^2)^2G^4=0.
\ee
An important observation is that in this case, as for $N_c=3$, the
boundary of the pion condensation phase is always a second order
transition, and there is no tricritical point. 
The complete phase diagram for $\mu_B=0$ is pictured in Fig.~\ref{fig4}. 
In the $\mu_I$-$T$--plane 
the critical endpoint of the first order chiral restoration
transition does not appear, because its position is in the region where
the pion condensation phase has a smaller free energy than the chiral
condensation phase.
We have shown  \cite{Klein:2003fy} that in this plane even for large
quark masses, the critical endpoint remains masked by the pion
condensation phase.

\subsubsection{The complete $\mu_I$-$\mu_B$--plane at $T\neq0$}
\setcounter{paragraph}{0}
To obtain the full picture of the phase diagram at nonzero values for
the two chemical potentials, the temperature and the quark mass, we
solve the saddle point equations % j4 for this case 
numerically. The
results for the phase diagram are presented in Fig.~\ref{fig5}.
It is in complete agreement with the expectations from our discussion
of the phase diagram at either $m=0$ or $T=0$: The finite quark mass
mainly affects the phase diagram at low values of the
chemical potentials, and the regions with either a diquark or a pion
condensate shrink with increasing temperature and finally vanish for a
temperature of order one, $TG={\mathcal O}(1)$.

We wish to draw the attention to one particular detail of the phase
diagram in our model.
~From Fig.~\ref{fig3}, 
the diquark and pion condensation phases are no longer contiguous at the
lines $\mu_1=0$ and $\mu_2=0$, and
there is no first order transition between them. Instead, this region
of the phase diagram is occupied once again by 
the chirally broken phases, and the chiral phase transition line reemerges. 
While the first order chiral transition line does not
appear for $\mu_B=0$ (and by symmetry not for $\mu_I=0$, either), we
thus find that for both $\mu_B$ and $\mu_I$ different from zero and an
appropriate value of the quark mass this first order line and its
critical endpoint can be observed. The phase diagram in the
$\mu_B$-$T$ plane at an isospin chemical potential that is chosen
to avoid the superfluid phases at low $\mu_B$ shares striking
similarities with the phase diagram of QCD with three colors at
nonzero baryon and isospin chemical potentials.

\subsubsection{The $\mu_B$-$T$--plane at constant $\mu_I$}

To study the chiral phase transitions for both $\mu_B$ and $\mu_I$
nonzero in some more detail, we turn to
the phase diagram in planes described by constant isospin
chemical potential.
Fig.~\ref{fig6} shows the phase diagram in these $\mu_B$-$T$--planes for fixed
values of $\mu_I$. The corresponding cuts through the phase diagram in
the three 
parameters $\mu_I$, $\mu_B$ and $T$ are perpendicular to the cuts at
constant $TG$ in Fig.~\ref{fig5}. Since
$\mu_1=\mu_B+\mu_I$ and $\mu_2=\mu_B-\mu_I$, planes at constant
$\mu_I$ intersect the planes with constant $T$ along lines
parallel to the diagonal from the lower  
left-hand to the upper right-hand corners of the panels in Fig.~\ref{fig5}.  

In our analysis of the plane $\mu_B=0$ and in
Fig.~\ref{fig4} above, we have seen that the first order
chiral phase transition and its 
critical endpoint are not present for any value of the
quark mass, and by virtue of the symmetries this also holds for the
plane given by $\mu_I=0$, shown in the first panel of
Fig.~\ref{fig6}. But with increasing value of the isospin chemical
potential the phase 
diagram in the $\mu_B$-$T$--plane changes considerably: first, the
first order chiral transition for $\sigma_1 \sim \langle \bar u
u\rangle$ at large temperature and its critical endpoint reappear.
With a further increase of $\mu_I$, the chiral transition for
$\sigma_2$ at large values of $\mu_B$ appears as well (see the second and
following panels of Fig.~\ref{fig6}). We have to caution, though, that
for the second chiral condensate $\sigma_2$ this is possible only
because of the 
saturation transition of the diquark condensate, which exists on the
lattice as well as in our random matrix model. If $\mu_I$ is made
larger still, the critical endpoint of the transition for $\sigma_1$
is once again swallowed by the pion condensation phase (see last panel
of Fig.~\ref{fig6}, where this is about to take place).

\begin{figure}
\hspace*{-0.4cm}
\includegraphics[scale=1.00, clip=true, angle=0,
draft=false]{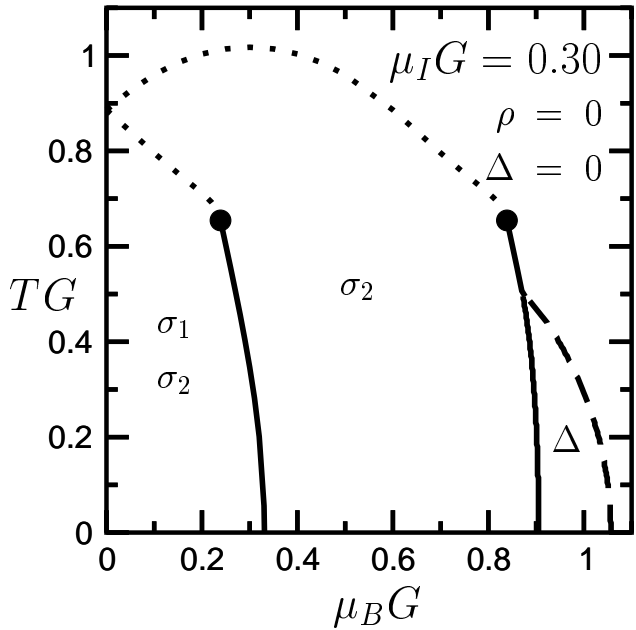} 
\caption{\label{fig7} Phase diagram in the $\mu_B$-$T$--plane at
  fixed values of the isospin chemical potential $\mu_I G=0.3$ 
for a finite quark mass of $mG=0.2$. First (second) order transitions are 
indicated by solid (dashed) lines, the dotted lines indicate a
  crossover. Phases are labeled by the non-vanishing 
condensates, or the condensates with large expectation values. This
phase diagram is very similar for either QCD with two colors or for
QCD with three colors.}
\end{figure}

For a quark mass $mG>0.184$ and at any temperature, the diquark and
pion superfluid phases are not contiguous. Therefore for a
sufficiently large quark mass ($mG>0.184$) and an 
isospin chemical potential that is
small enough so that the pion superfluid phase is never reached
and large enough so that the diquark superfluid phase is reached only
at large $\mu_B$, the structure of the phase diagram is as in 
Fig.~\ref{fig7}.  
 The resulting phase diagram is strikingly similar to
the phase diagram of the   Random Matrix Model for QCD with three
color at nonzero baryon and isospin chemical potentials
\cite{Klein:2003fy}. 

%%%%%%%%%%%%%%%%%%%%%%%%%%%%%%%%%%%%%%%%%%%%%%%%%%%%%%%%%%%%%%%%%%%%%%%%%%%
%%%%%%%%%%%%%%%%%%%%%%%%%%%%%%%%%%%%%%%%%%%%%%%%%%%%%%%%%%%%%%%%%%%%%%%%%%% 

\section{Conclusion} 
\label{nc2section-6} 
\setcounter{equation}{0} 
In this article, we have introduced a random matrix model with the
symmetries of QCD with two colors and two flavors and nonzero
baryon chemical potential, isospin chemical potential and temperature. 
Contrary to work elsewhere in the literature, our temperature term
respects the flavor symmetry of the partition function as it should.
 
The random matrix partition function has been 
rewritten exactly in terms of an effective Lagrangian.
To lowest order in the chiral perturbation expansion, the effective Lagrangian 
agrees with the one derived in chiral
perturbation theory \cite{Kogut:1999iv, Kogut:2000ek,
  Splittorff:2000mm}.
We have analyzed the phase diagram of this model 
for values of the mass and chemical potential outside of the domain
of chiral perturbation theory. 
It shows a rich phase structure 
with phases
characterized by the diquark condensate, the pion condensate and the chiral
condensates, $\langle\bar u u\rangle$ and $\langle \bar d d \rangle$.
In the presence of two chemical potentials,
the chiral condensates are not necessarily equal. For a nonzero quark
mass, in regions
without pion or baryonic condensate at small values for the
chemical potentials, the restoration of the spontaneously broken
symmetry takes place via two separate crossovers, across which the
chiral condensates for the two flavors rapidly change in value.  
The vanishing of the superfluid phases at large values of
the relevant chemical potentials, which is also observed on the
lattice, has been explained as a saturation effect.
On the lattice, this is due to the finite number of lattice sites
\cite{Kogut:2001na}, and for random matrix models, this is   
due the compactness of the Dirac operator.

The tricritical point of the diquark
and pion condensation transitions observed in lattice calculations
\cite{Kogut:2001if, Kogut:2002tm, Kogut:2002zg, Kogut:2002cm} is
absent in the random matrix models. 
Since the Random Matrix partition
function is equivalent to a mean field description of the phase
transition, this is not entirely surprising. Symmetries alone
do not determine the presence of a tricritical point.
In chiral perturbation
theory, a tricritical point is obtained from a one-loop effective
potential \cite{Splittorff:2002xn,dunne}.
Ultimately, the question of the
structure of the phase diagram in this regard must be answered by
lattice calculations \cite{Kogut:2002cm}.

Finally, we would like to compare our results with the 
random matrix study of the
phase diagram for three-color QCD at nonzero temperature, baryon and
isospin chemical potentials \cite{Klein:2003fy}. 
First we notice that the upper-left quadrant of
the phase diagram in the $\mu_B$-$\mu_I$--plane is identical in both
cases. This is due to the fact that in both cases, the isospin
chemical potential is related to Goldstone bosons of the microscopic
theory. The upper-right quadrants are obviously different in the two
theories, since there
is no Goldstone boson that carries baryon number in three-color
QCD. However, the phase diagram for three-color QCD with three quark
flavors of equal masses at nonzero isospin and strange chemical
potential should be very similar to the one we have presented here. 

Second, and more importantly, we notice that it is possible to find
a quark mass and an isospin chemical potential for which the diquark
and pion condensation phases appear only in certain regions of the
phase diagram. Namely, they are present only at low temperature and
either very small or 
very large values of the baryon chemical potential. As shown in
Figs.~\ref{fig6} and \ref{fig7}, we see that the first order
transition lines of the 
chiral transitions and their critical endpoints are not at all affected
by the phases with bosonic condensates. Indeed, the transition lines
are identical to those that are found in the random matrix model for
three-color QCD. This observation implies that the critical endpoint
of three-color QCD can be studied in two-color QCD. 
For an appropriate choice of quark mass and isospin chemical potential,
the phase diagram in the $\mu_B$-$T$ plane are almost identical for
QCD with two colors and for QCD with three colors.
In particular, the doubling of the critical lines due to a nonzero
isospin chemical potential can be studied in QCD
with two colors.

\begin{acknowledgments}
J. Kogut, K. Splittorff and B. Vanderheyden are acknowledged for
useful discussions. D.T. is supported in part by the
``Holderbank''-Stiftung. This work was partially supported by the
U.S. DOE Grant No. DE-FG-88ER40388 and by the NSF under Grant
No. NSF-PHY-0102409.  
\end{acknowledgments}

% thebibliography is in file nc2bib.tex

\end{document}